\documentclass[11pt]{article}

\usepackage{mymacros}

\title{Black holes and quantum entanglement}

\author[]{Ro Jefferson}
\affiliation[]{Max Planck Institute for Gravitational Physics (Albert Einstein Institute),\\Am M\"uhlenberg 1, D-14476 Potsdam-Golm, Germany}
\emailAdd{rjefferson@aei.mpg.de}

\abstract{
The black hole information paradox is a contradiction between fundamental principles which has puzzled physicists for over forty years. The crux of the problem lies in an assumption about the structure of entanglement across the event horizon, namely, that the Hilbert space factorizes. While valid in quantum mechanics, this fails drastically in quantum field theory, and hence a deeper understanding of entanglement is required if further progress is to be made. Recently, ideas from algebraic quantum field theory have provided new insight into this issue, and show promise for elucidating the connection between entropy and horizons that underlies black hole thermodynamics.
\vfill
\centerline{\emph{Essay written for the Black Hole Initiative's 2018 paper competition.}}
}

\begin{document}
\maketitle

In 1973, Jacob Bekenstein observed \cite{PhysRevD.7.2333} that black holes must be endowed with an entropy in order to preserve the second law of thermodynamics; otherwise, one could decrease the entropy of the universe by simply throwing subsystems with high entropy into a black hole. At face value, this is an intuitive proposal: since the information about the degrees of freedom that comprise the hypothetical subsystem would then be hidden behind the event horizon, it makes sense to count them among the microstates of the black hole.

The unintuitive twist comes from the realization that this na\"ive bookkeeping is not at all how black holes operate. The entropy of familiar systems scales with the volume thereof, which is consistent with simply counting the constituent (e.g., particulate) degrees of freedom. Black hole entropy, in stark contrast, scales with the \emph{area} of the event horizon. Bekenstein's original motivation for this claim hinged largely on Stephen Hawking's 1971 result \cite{PhysRevLett.26.1344} that the surface area of a black hole cannot decrease in any classical process (the so-called ``area theorem''). This lead Bekenstein to propose an analogy between black holes and statistical thermodynamics, which has since been enshrined in the laws of \emph{black hole thermodynamics}.

Classically however, black holes do not radiate (hence the name), and therefore have zero temperature and consequently zero thermodynamic entropy. For this reason, black hole thermodynamics initially appeared purely epistemic---a reflection of our knowledge of reality, rather than the real (or ontic) nature thereof. But this metaphysical state of affairs soon changed when Hawking showed \cite{Hawking1975} that, quantum mechanically, black holes \emph{do} radiate, with characteristic temperature
\be
T=\frac{1}{8\pi M}~,
\ee
and entropy
\be
S=\frac{A}{4}~,\label{eq:S}
\ee
where $M$ is the mass of the black hole, and $A$ is the area of its event horizon. The emission of this \emph{Hawking radiation} implies that black holes can evaporate, and thus their surface area $A$, and by extension their entropy $S$, can in fact decrease when quantum effects are taken into account. Fortuitously, including the entropy of the Hawking radiation in the total more than suffices to compensate for this decrease: the second law of thermodynamics is saved.

With Hawking's discovery that black holes are not completely black after all, black hole thermodynamics went from epistemic to ontic in one fell swoop---and brought shattering implications in its wake. By far the most notorious is the \emph{information paradox} \cite{PhysRevD.14.2460}, which refers to the apparent loss of information that results from a black hole which forms from a pure state, but upon evaporating leaves nothing but a mixed state of thermal radiation. This represents a grievous violation of quantum mechanics---specifically, the principle of unitarity, which ensures that probabilities are conserved. In a nutshell, quantum mechanics says that information can never be truly lost. Black holes appear to disagree---strenuously.

This is a deeply disturbing result: it implies a contradiction between the fundamental pillars on which the whole edifice of modern physics stands. And despite nearly 45 years of intensive effort, the solution continues to elude us. Indeed, the enduring severity of the conflict is aptly illustrated by the recent \emph{firewall} controversy, which was ignited by a 2013 paper \cite{Almheiri:2012rt} by Ahmed Almheiri, Donald Marolf, Joseph Polchinski, and James Sully (AMPS). In it, the authors honed the long-standing information paradox to razor sharpness, and argued that giving up Einstein's equivalence principle -- which underlies general relativity -- is the most conservative resolution. But in addition to noticeable violations of low-energy physics, this would imply that an intrepid explorer, rather than falling smoothly through the event horizon as Einstein predicts, instead gets incinerated by a ``firewall'' of ultra high-energy quanta. Clearly, general relativity cannot be so easily tampered with!

It is illuminating to contrast black hole evaporation with the apparently pure-to-thermal evolution of normal matter upon incineration, say a burning lump of coal. Supposing this to be in an initially pure state, the final state again involves a thermal bath of radiation, with the apparent loss of information that implies. But physicists are not concerned about unitarity-violating barbecues. The reason is that subtle correlations between early and late radiation conspire to preserve the purity of the total system. It is only due to our coarse-grained, inevitably imperfect measurements that we perceive a thermal state. In other words, while it may be impossible to actually recover the information in practice, in principle a sufficiently powerful computer could do it. The laws of quantum mechanics survive unscathed.

The crucial difference between the coal and the black hole is that the former has no event horizon. Early quanta are entangled with quanta inside the coal, which can -- via their interactions with other interior quanta -- imprint information on the late radiation. In contrast, the presence of a horizon imposes a very specific, \emph{pairwise entanglement} structure on the Hawking quanta across it, which forbids them from sharing their entanglement as in the lump of coal. \emph{It is this quantum structure at the horizon that ultimately underlies the black hole information paradox}. Indeed, this entanglement structure was used explicitly by AMPS when they kindled the firewall, and has been implicitly assumed in nearly every discussion of Hawking radiation to date. 

However, while the decomposition of the vacuum that underlies this pairwise entanglement is perfectly valid in quantum mechanics, it fails drastically in quantum field theory. The problem is that Hilbert space -- the abstract space where quantum states live -- does not factorize into subspaces (in this case, the inside and outside of the black hole). This manifests in the fact that the entanglement entropy between two regions in quantum field theory is formally infinite. In other words, the vacuum is an infinitely entangled state, and this fact has consequences which we're only beginning to understand.

Perhaps the most famous example of this is the Reeh-Schlieder theorem, which uses the infinite entanglement of the vacuum to shake our notion of locality to the core. Na\"ively, one might suppose that a localized subregion contains no information about the universe beyond. Reeh-Shlieder proves otherwise: there exist operators confined to such regions, which nonetheless allow one to reconstruct any other state in the global Hilbert space with arbitrary accuracy. To illustrate how shocking this is to theorists, it means that by acting with such an operator in the room in which you're reading this, you could create the Moon---or the supermassive black hole at the center of our galaxy!

This is more than merely a mathematical curiosity. Rather, it hints at an incredible richness to which the quantum mechanical approach to entanglement entropy, with its simplistic assumption about Hilbert space factorization, is blind. And the lesson from black holes is that we must mine this deeper wealth in order to escape from the paradoxes in which four and half decades of inquiry from faulty premises have mired us. Black holes force us to rethink everything we think we know about entanglement in quantum field theory, and to develop mathematical tools capable of casting light where our meagre intuition cannot tread. 

In fact, there exists a mathematical framework which is particularly well-adapted to foundational questions in both physics and the philosophy thereof, known as algebraic quantum field theory (AQFT). This was developed starting in the 1950's as a mathematically rigorous approach to quantum field theory, but has historically been a relatively esoteric pursuit of mathematical physicists, of little practical use for calculating anything one might wish to measure in a particle accelerator. But this abstract formalism also enjoys powerful advantages; in particular, the pathological divergences that plague the standard approach to entanglement above simply do not arise. For this reason, ideas from AQFT have now begun to surface in the theoretical physics community as promising new tools to deepen our understanding of entanglement \cite{Witten:2018zxz}. For example, they have been used to shed light on aspects of locality and causality in the context of the holographic principle (itself inspired by black hole thermodynamics), thereby further elucidating the emergent relationship between entanglement and spacetime geometry \cite{Faulkner:2013ica,Faulkner:2017vdd}. More pertinently, this framework has recently been applied to black hole interiors \cite{Papadodimas:2017qit,Jefferson:2018ksk}, in an effort to understand how this information might be represented by an external observer.

The nascent application of AQFT to the study of black holes represents a novel and interdisciplinary approach, which draws on the combined strength of physics, mathematics, and philosophy to extract insights about quantum entanglement and the structure of spacetime itself. At the very least, it hints at a way beyond the impasse created by the superficial treatment of entanglement above, and may elucidate the deeper connection between entropy and horizons that underlies black hole thermodynamics. And while it is too soon to tell whether this approach will ultimately succeed in finally extinguishing the firewall paradox, it is already clear that there are many fascinating insights about black holes yet to be gained.

\pagebreak

\bibliographystyle{ytphys}
\bibliography{biblio}

\providecommand{\href}[2]{#2}\begingroup\raggedright\begin{thebibliography}{10}

\bibitem{PhysRevD.7.2333}
J.~D. Bekenstein, ``Black holes and entropy,''
  \href{https://link.aps.org/doi/10.1103/PhysRevD.7.2333}{{\em Phys. Rev. D}
  {\bfseries 7} (Apr, 1973) 2333--2346}.

\bibitem{PhysRevLett.26.1344}
S.~W. Hawking, ``Gravitational radiation from colliding black holes,''
  \href{https://link.aps.org/doi/10.1103/PhysRevLett.26.1344}{{\em Phys. Rev.
  Lett.} {\bfseries 26} (May, 1971) 1344--1346}.

\bibitem{Hawking1975}
S.~W. Hawking, ``Particle creation by black holes,''
  \href{http://projecteuclid.org/euclid.cmp/1103899181}{{\em Comm. Math. Phys.}
  {\bfseries 43} no.~3, (1975) 199--220}.

\bibitem{PhysRevD.14.2460}
S.~W. Hawking, ``Breakdown of predictability in gravitational collapse,''
  \href{https://link.aps.org/doi/10.1103/PhysRevD.14.2460}{{\em Phys. Rev. D}
  {\bfseries 14} (Nov, 1976) 2460--2473}.

\bibitem{Almheiri:2012rt}
A.~Almheiri, D.~Marolf, J.~Polchinski, and J.~Sully, ``{Black Holes:
  Complementarity or Firewalls?},''
  \href{http://dx.doi.org/10.1007/JHEP02(2013)062}{{\em JHEP} {\bfseries 02}
  (2013) 062},
\href{http://arxiv.org/abs/1207.3123}{{\ttfamily arXiv:1207.3123 [hep-th]}}.

\bibitem{Witten:2018zxz}
E.~Witten, ``{Notes on Some Entanglement Properties of Quantum Field Theory},''
\href{http://arxiv.org/abs/1803.04993}{{\ttfamily arXiv:1803.04993 [hep-th]}}.

\bibitem{Faulkner:2013ica}
T.~Faulkner, M.~Guica, T.~Hartman, R.~C. Myers, and M.~Van~Raamsdonk,
  ``{Gravitation from Entanglement in Holographic CFTs},''
  \href{http://dx.doi.org/10.1007/JHEP03(2014)051}{{\em JHEP} {\bfseries 03}
  (2014) 051},
\href{http://arxiv.org/abs/1312.7856}{{\ttfamily arXiv:1312.7856 [hep-th]}}.

\bibitem{Faulkner:2017vdd}
T.~Faulkner and A.~Lewkowycz, ``{Bulk locality from modular flow},''
  \href{http://dx.doi.org/10.1007/JHEP07(2017)151}{{\em JHEP} {\bfseries 07}
  (2017) 151},
\href{http://arxiv.org/abs/1704.05464}{{\ttfamily arXiv:1704.05464 [hep-th]}}.

\bibitem{Papadodimas:2017qit}
K.~Papadodimas, ``{A class of non-equilibrium states and the black hole
  interior},''
\href{http://arxiv.org/abs/1708.06328}{{\ttfamily arXiv:1708.06328 [hep-th]}}.

\bibitem{Jefferson:2018ksk}
R.~Jefferson, ``{Comments on black hole interiors and modular inclusions},''
\href{http://arxiv.org/abs/1811.08900}{{\ttfamily arXiv:1811.08900 [hep-th]}}.

\end{thebibliography}\endgroup
\end{document}